\documentclass[doublecol]{epl2}
 \newcommand{\be}{\begin{equation}}
 \newcommand{\ee}{\end{equation}}
 \newcommand{\ben}{\begin{eqnarray}}
 \newcommand{\een}{\end{eqnarray}}
 
  \usepackage[dvipdf]{epsfig}
\usepackage{color}

\title{ Fermionic cosmologies with Yukawa type interactions}
\shorttitle{ Fermionic cosmologies with Yukawa type interactions}
\author{Marlos O. Ribas\inst{1}\footnote{gravitam@yahoo.com} \and Fernando P. Devecchi\inst{2}\footnote{devecchi@fisica.ufpr.br} \and Gilberto M. Kremer\inst{2}\footnote{kremer@fisica.ufpr.br}}
\shortauthor{M. O. Ribas \etal}
\institute{
  \inst{1} Departamento de F\'{\i}sica, Universidade Tecnol\'ogica Federal do Paran\'a,
 Curitiba, Brazil\\
  \inst{2} Departamento de F\'{\i}sica,
Universidade Federal do Paran\'a, Curitiba, Brazil
}

\abstract{
In this work we discuss  if  fermionic sources could be
responsible for accelerated periods in a Friedmann-Robertson-Walker spatially flat universe, including  a usual self-interaction potential of the Nambu-Jona-Lasinio type  together with a fermion-scalar interaction potential  of the Yukawa type. The results show that the combination of these  potentials could promote an initially accelerated period, going through a middle decelerated era, with a final eternal accelerated period, where the self-interaction contribution dominates.}


\begin{document}

\maketitle

The identification of  constituents that promoted the inflationary   period in
the evolution of the universe is a fundamental topic in cosmology. Several candidates has been
tested for describing both the inflationary period and the present
accelerated era: scalar fields, exotic equations of state and
cosmological constant \cite{Peeb}. Besides those,  fermionic fields has also been  tested as gravitational
sources of an expanding universe.   In fact, fermionic sources can be  responsible of  accelerated
periods with different regimes emerging from it \cite{Fermions,RDK,Saha,Armendariz}. In
some of these  models the fermionic field plays the role of the inflaton in the
early period of the universe and of dark energy for the old
universe, without the need of a cosmological constant term or a
scalar field. In an old universe scenario an initially matter
dominated period gradually turns into a dark (fermion) energy
period when an accelerated regime starts and remains for the rest of
the evolution of the system. These fermionic sources has been
investigated using several
 approaches, with results including numerical solutions, exact solutions,
anisotropy-to-isotropy scenarios and cyclic cosmologies (see, for
example~\cite{Fermions,RDK,Saha,Armendariz}). When considering  these models, one important
point is the choice of the fermionic potential and in previous works  \cite{RDK} self-interacting potentials were tested to account for accelerated regimes.
One complementary/alternative approach  would be consider interactions between a scalar and a fermionic field through the presence of a  potential of the Yukawa type
\cite{Ryder}. This potential was proposed originally in particle physics to describe the behavior of strong force interactive fermions and
 here we   want to test which kind of role  the Yukawa  potential could play in an young accelerating universe.
Besides, as in previous works \cite{RDK,Saha}, a fermionic self-interaction  term  of the Nambu-Jona-Lasinio type  \cite{NJL} is included.
We have used the metric signature $(+,-,-,-)$ and natural units with $8\pi G=c=\hbar=1$.

Let us start with  a brief review of the elements of the tetrad formalism employed  to include fermionic fields in a dynamical curved space-time, since,
 as it is well known, the tetrad formalism permits the inclusion of fermions in gravitational models. Following
the general covariance principle, a connection between the tetrad
and the metric tensor $g_{\mu\nu}$ is established through the
relation
$ g_{\mu\gamma} =e^a_\mu e^b_\gamma\eta_{ab},  a=0,1,2,3 $ where $e^a_\mu $ denotes the tetrad or   vierbein and $\eta_{ab}$ is the Minkowski
metric tensor (see e.g. \cite{Wald,Ryder,Weinberg}). Here Latin indices refer to the  local  inertial frame  whereas Greek indices to the  general system.
Furthermore, the general covariance principle
imposes that the usual Dirac-Pauli matrices $\gamma^a$ must be replaced by
their generalized counterparts $\Gamma ^{\mu}=e^\mu_a\gamma^a$,
where these   matrices satisfy the  extended
Clifford algebra, i.e., $ \{\Gamma^\mu,\Gamma^\nu\}=2g^{\mu\nu}.$

The generally covariant Dirac Lagrangian density for the fermionic field  is given by
\begin{equation}
\mathcal{L}_{\rm Dirac}(\psi)=\frac{\imath}{2}[ \overline\psi\,\Gamma^\mu
D_\mu\psi-(D_\mu\overline\psi)\Gamma^\mu\psi]-V(\psi),
\label{5}
\end{equation}
where $\psi$ and $\overline\psi=\psi^\dag\gamma^0$ denote the spinor field and its adjoint, respectively, and $V(\psi)$ is the self-interaction potential  of the fermionic field which is supposed  to be massless. In the above equation the covariant derivatives read
\begin{equation}
D_\mu\psi= \partial_\mu\psi-\Omega_\mu\psi,\qquad
D_\mu\overline\psi=\partial_\mu\overline\psi+\overline\psi\Omega_\mu,
 \label{3}
\end{equation}
where  the spin connection $\Omega_\mu$ is given by
\begin{equation}
\Omega_\mu=-\frac{1}{4}g_{\rho\sigma}[\Gamma^\rho_{\mu\delta}
-e_b^\rho(\partial_\mu e_\delta^b)]\Gamma^\delta\Gamma^\sigma,
\label{4}
\end{equation}
with $\Gamma^\nu_{\sigma\lambda}$ denoting the Christoffel symbols.

The Lagrangian density for a massive scalar field $\phi$ and the one corresponding to the Yukawa interaction between  the fermionic and the scalar field read
\ben
&&\mathcal{L}_{\rm scalar}(\phi)={1\over2}\partial^\mu\phi\,\partial_\mu\phi-{1\over2}m^2\phi^2,\\
&&\mathcal{L}_{\rm Yukawa}(\phi,\psi)=-\lambda\overline\psi\phi\psi,
\een
where $m$ is the mass of the scalar field and $\lambda$ the coupling constant of the Yukawa potential.

The total action for a massless fermionic field  and a massive scalar field that are connected through an interaction of Yukawa type can be written as
\begin{eqnarray}\nonumber
S=\int\sqrt{-g}d^4x\Bigg\{\frac{1}{2}R+{1\over2}\partial^\mu\phi\,\partial_\mu\phi-{1\over2}m^2\phi^2-\lambda\overline\psi\phi\psi
\\\label{5b}
+\frac{\imath}{2}\left[\overline\psi\Gamma^\mu D_\mu\psi-(D_\mu\overline\psi)\Gamma^\mu\psi\right]-V(\psi)\Bigg\},\qquad
\end{eqnarray}
where $R$ is the scalar curvature.

 We  obtain the Dirac equations for the spinor field  and its adjoint coupled to the gravitational field from the Euler-Lagrange equations for $\psi$ and $\overline\psi$,  namely
\ben\label{7a}
\imath\Gamma^\mu D_\mu\psi-{\partial V\over \partial{\overline\psi}}-\lambda\phi\psi= 0,\\\label{7b}
\imath D_\mu\overline\psi\,\Gamma^\mu+{\partial V\over \partial\psi}+\lambda\overline\psi\phi= 0.
\een

Likewise the Euler-Lagrange equation for $\phi$ leads to the modified Klein-Gordon equation
\begin{equation}\label{7c}
\nabla_\mu\nabla^\mu\phi+m^2\phi+\lambda\overline\psi\psi=0.
\end{equation}

 From the variation of the total action (\ref{5b})  with respect to the tetrad we obtain Einstein field equations
\begin{equation}\label{7d}
R_{\mu\nu}-\frac{1}{2}g_{\mu\nu}R=-T_{\mu\nu},\label{8}
\end{equation}
 where  $T_{\mu\nu} $ is the total energy-momentum tensor which
 is a sum of the contributions from the fermionic and scalar fields. Since we are dealing with a fermionic field in a space-time without torsion, the
total energy-momentum tensor is symmetric and reads
\begin{eqnarray}\nonumber
T^{\mu\nu}=\frac{\imath}{4}\left[\overline\psi\Gamma^\mu D^\nu\psi+\overline\psi\Gamma^\nu D^\mu\psi-D^\nu\overline\psi\Gamma^\mu\psi \right.
\\\left.-D^\mu\overline\psi\Gamma^\nu\psi\right] +\partial^\mu\phi\partial^\nu\phi-g^{\mu\nu}\left[{1\over2}\partial^\sigma\phi\,\partial_\sigma\phi-{1\over2}m^2\phi^2\right. \nonumber \\\left.
-\lambda\overline\psi\phi\psi+\frac{\imath}{2}\left(\overline\psi\Gamma^\lambda D_\lambda\psi-D_\lambda\overline\psi\Gamma^\lambda\psi\right)-V(\psi)\right].
\end{eqnarray}

 The  Friedmann-Robertson-Walker metric
 \begin{equation}
ds^2=dt^2-a(t)^2(dx^2+dy^2+dz^2), \label{11}
\end{equation}
with $a(t)$ denoting the cosmic scale factor, mirrors the homogeneity and isotropy properties of a spatially flat universe. For this metric the components of the tetrad, Dirac-Pauli matrices and spin connection  read
\ben
e_0^\mu=\delta_0^\mu,\qquad e_i^\mu=\frac{1}{a(t)}\delta_i^\mu, \qquad
\Gamma^0=\gamma^0, \\
\Gamma^i=\frac{1}{a(t)}\gamma^i,\qquad\Omega_0=0,\qquad  \Omega_i=\frac{1}{2}\dot
a(t)\gamma^i\gamma^0,
\een
where the dot represents the derivative with respect to time.

Furthermore, from the hypothesis of homogeneity and isotropy it follows that the spinor and scalar fields depend only on time so  that the Dirac (\ref{7a}), (\ref{7b}) and Klein-Gordon
(\ref{7c}) equations reduce to
\ben\label{8a}
\dot\psi+\frac{3}{2}\frac{\dot a}{a}\psi+\imath\gamma^0{\partial V\over \partial{\overline\psi}}+\imath \lambda\gamma^0\psi\phi=0,\\\label{8b}
\dot{\overline\psi}+\frac{3}{2}\frac{\dot a}{a}\overline\psi-\imath{\partial V\over \partial\psi}\gamma^0-\imath\lambda\phi\overline\psi\gamma^0=0,\\\label{8c}
\ddot\phi+3\frac{\dot a}{a}\dot\phi+m^2\phi+\lambda\overline\psi\psi=0.
\een

From the Einstein field equations  (\ref{7d}) it follows the
 Friedmann and acceleration equations
 \be\label{9}
 \left({\dot a\over a}\right)^2={1\over3}\rho,\qquad {\ddot a\over a}=-{1\over 6}(\rho+3p),
 \ee
respectively, where the total energy density $\rho$ and total pressure $p$ are  are given by
\ben\label{10a}
&&\rho=\frac{1}{2}\dot\phi^2+\frac{1}{2}m^2\phi^2+\lambda \phi(\overline\psi\psi)+V(\psi),\\\label{10b}
&&p=\frac{1}{2}\dot\phi^2-\frac{1}{2}m^2\phi^2-V(\psi)+{\partial V\over\partial\psi}{\psi\over2}+{\overline\psi\over2}{\partial V\over\partial\overline\psi}.\qquad
\een

We suppose now that the self-interaction potential of the fermionic field is given by $V=\Lambda\left(\overline\psi\psi\right)^n$, where $\Lambda$ is a coupling constant and $n$  a constant exponent. If we substitute this potential into the Dirac equations (\ref{8a}) and (\ref{8b}), multiply the first one by $\overline\psi$, the second one by $\psi$ and add the resulting equations we obtain a differential equation for the bilinear $\overline\psi\psi$ whose solution is given by $\overline\psi\psi=C/a^3$ where $C$ denotes an integration constant.

Once we know the explicit form of the bilinear $\overline\psi\psi=C/a^3$ the total energy density (\ref{10a}) and total pressure (\ref{10b}) become
\ben\label{11a}
&&\rho=\frac{1}{2}\dot\phi^2+\frac{1}{2}m^2\phi^2+ {C_1\over a^3}\phi+ {C_2\over a^{3n}},\\\label{11b}
&&p=\frac{1}{2}\dot\phi^2-\frac{1}{2}m^2\phi^2+(n-1) {C_2\over a^{3n}},
\een
where we have introduce the constants $C_1=C\lambda$ and $C_2=C\Lambda$.

Therefore, from (\ref{8c}) and (\ref{9})$_2$ we have a coupled system of non-linear differential equations for the determination of $a(t)$ and $\phi(t)$ that include the modified Klein-Gordon equation and the acceleration equation, namely,
\ben\label{12a}
\ddot\phi&+&3\frac{\dot a}{a}\dot\phi+m^2\phi+{C_1\over a^3}=0,\\\label{12b}
{\ddot a\over a}&+&{1\over 6}\left[2\dot\phi^2-m^2\phi^2+{C_1\over a^3}\phi+ (3n-2){C_2\over a^{3n}}\right]=0.\qquad
\een

The coupled system of differential equations (\ref{12a}) and (\ref{12b}) was solved numerically with the initial conditions that the cosmic scale factor and the energy density were normalized at $t=0$, i.e., $a(0)=1$ and $\rho(0)=1$, which from the Friedmann equation (\ref{9})$_1$ implies that $\dot a(0)=1/\sqrt{3}$. Furthermore, it was supposed that the scalar field at time $t=0$ was zero with a very small slope, i.e., $\phi(0)=0$ and $\dot\phi(0)=10^{-4}$. With these initial conditions, the coupling constant related with the self-interaction potential of the fermionic field is no more arbitrary,  since from the expression for the energy density (\ref{11a}) we must have $C_2=\rho(0)-\dot\phi(0)^2/2$.
However, it remains to specify the coupling constant of the Yukawa potential $C_1$, the mass of the scalar field $m$ and the exponent of the self-interaction potential of the fermionic field $n$. We have chosen fixed normalized values for  the mass of the scalar field and  for the exponent of the self-interaction potential of the scalar field -- namely, $m=10^{-6}$ and $n=0.5$ -- and variable values for the coupling constant of the Yukawa potential in order to see the influence of the strength of this potential on the solutions of the coupled system of differential equations.

\begin{figure}
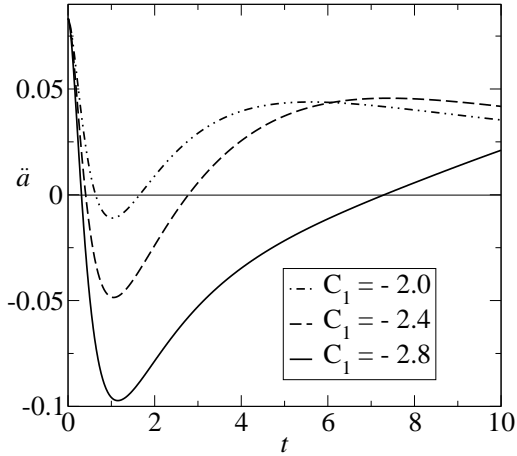

\onefigure[height=6cm]{fig1.eps}
\caption{Acceleration field $\ddot a$ as a function of time $t$.}
\label{fig.1}
\end{figure}

As a result of the numerical integration, the scale factor behaves accordingly to an ever-expansive universe.
The classification of eras can be done in terms of the time evolution of the acceleration field. In fact,
in Figure 1 we have plotted the behavior of the acceleration
 $\ddot a$ as a function of time, for three different (normalized) values of the  coupling constant of the Yukawa potential: $\vert C_1\vert=2.0,$ $2.4$ and $2.8$.
From this figure it is possible to infer that we have three periods which can be interpreted as an initial inflationary period where the acceleration is positive, a period dominated by radiation/matter where the acceleration is negative and a period dominated by the dark energy where the acceleration becomes again positive. Furthermore, we observe that longer  periods of deceleration are related with larger  values of $\vert C_1\vert$. By increasing the value of $\vert C_1\vert$ beyond $2.8$ the solution of acceleration field presents only an accelerated period followed by a decelerated one, while for values of $\vert C_1\vert$ smaller than $2.0$ the acceleration field has only one solution corresponding to an accelerated period.

In Figure 2 we show the behavior of the Yukawa potential $\overline\psi\phi\psi$ as function of time for different values of the coupling constant $C_1$. The behavior of the curves in this figure corroborates with the explanation given above, i.e., by increase of the coupling constant the amplitude of Yukawa potential becomes more accentuated, so that it has a direct influence on the decelerated period.

 The coupling constant of the Yukawa potential has also an influence in the time decay of the total energy density. This fact can be observed from the analysis of Figure 3 where the total energy density is plotted as a function of time for different values of $C_1$.

\begin{figure}
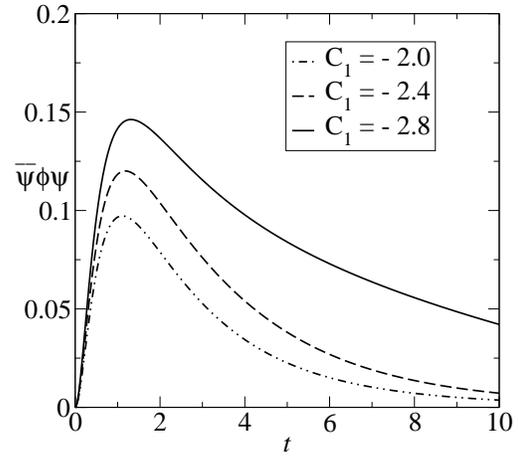

\onefigure[height=6cm]{fig2.eps}
\caption{Yukawa potential $\overline\psi\phi\psi$ as a function of time $t$.}
\label{fig.2}
\end{figure}

To sum up, in this work we have investigated the possibility of a fermionic cosmology to describe the different accelerated regimes of our universe.
This is possible due to the inclusion of two potentials, one self-interactive and one of the Yukawa type. When the universe is still young, the
combination of these contributions promote a short positive accelerated regime that can be associated to a inflationary period, followed by a decelerated period
that would correspond to a matter/radiation dominated universe.   We conclude that the intensity of the Yukawa coupling determines the time interval and the
amplitude of the decelerated regime. On the other hand, the constant that controls the intensity of the self-interactive potential is decisive to the existence of
a final accelerated period, that could correspond to a universe dominated by dark energy.

\begin{figure}
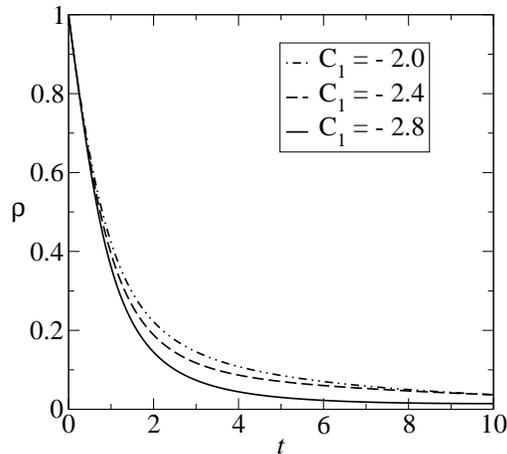

\onefigure[height=6cm]{fig3.eps}
\caption{Total energy density  $\rho$ as a function of time $t$.}
\label{fig.3}
\end{figure}

\acknowledgments
{ FPD and GMK acknowledge the
support by
Conselho Nacional de Desenvolvimento Cient\'{\i}fico e Tecnol\'ogico (CNPq). }

\end{document}